\documentstyle{article}

\def\be{\begin{equation}}
\def\ee{\end{equation}}
\def\bea{\begin{eqnarray}}
\def\eea{\end{eqnarray}}
\begin{document}

\pagestyle{empty}
\vskip-10pt
\hfill {\tt hep-th/0006231}

\hfill {revised October 2000}
\begin{center}
\vskip 5truecm
{\LARGE \bf
A class of six-dimensional conformal field theories
}\\ 
\vskip 2truecm
{\large\bf
M{\aa}ns Henningson
}\\
\vskip 1truecm
{\it Institute of Theoretical  Physics, Chalmers University of Technology\\
S-412 96 G\"{o}teborg, Sweden}\\
\vskip 5truemm
{\tt mans@fy.chalmers.se}
\end{center}
\vskip 2truecm
\noindent{\bf Abstract:}
We describe a class of six-dimensional conformal field theories that have some properties in common with and possibly are related to a subsector of the tensionless string theories. The latter theories can for example give rise to four-dimensional $N = 4$ superconformal Yang-Mills theories upon compactification on a two-torus. Just like the tensionless string theories, our theories have an $ADE$-classification, but no other discrete or continuous parameters. The Hilbert space carries an irreducible representation of the same Heisenberg group that appears in the tensionless string theories, and the `Wilson surface' observables obey the same superselection rules. When compactified on a two-torus, they have the same behaviour under $S$-duality as super Yang-Mills theory.  Our theories are natural generalizations of the two-form with self-dual field strength that is part of the world-volume theory of a single five-brane in $M$-theory, and the $A_{N - 1}$ theory can in fact be seen as arising from $N$ non-interacting chiral two-forms by factoring out the collective `center of mass' degrees of freedom.   

\newpage
\pagestyle{plain}

\section{Introduction}
Among the most surprising discoveries of recent years is the existence of consistent theories in six Minkowski dimensions that do not include dynamical gravity. The IR-limit of these so called tensionless string theories are $(0, 2)$ superconformal field theories that obey an $ADE$-classification, but have no other discrete or continuous parameters. They can be realized by considering type IIB string theory on a six-manifold $M$ times a hyper-K\"ahler four-manifold that develops the corresponding $ADE$-type singularity \cite{9507121}. At the locus of the singularity, there is then a six-dimensional conformal field theory that decouples from the bulk degrees of freedom. Type IIB three-branes that wrap on vanishing two-spheres, corresponding to the roots of the $ADE$-type Lie algebra, are seen as tensionless strings in six dimensions. The $A_{N - 1}$ theory can also be realized as the world-volume theory of $N$ coincident five-branes in $M$-theory \cite{9512059}. The tensionless strings then arise from membranes stretching between different five-branes, corresponding to the roots of the $A_{N-1}$ Lie algebra. The $A_{N - 1}$ theory can of course be regarded as a natural generalization of the `$A_0$' theory defined on the world-volume of a single five-brane \cite{9510053}. This is the theory of a free $(0, 2)$ supersymmetric tensor multiplet, comprising a two-form `gauge field' $X$ with self-dual field strength, i.e. $d X = {}^* d X$, a set of scalars $\phi^i$ transforming in the $\bf 5$ representation of the $SO (5) \simeq Sp (4)$ $R$-symmetry group, and a symplectic Majorana-Weyl fermion $\psi^a$ transforming in the $\bf 4$ representation.

The tensionless string theories are in many ways analogous to $N = 4$ super Yang-Mills theory with an $ADE$ gauge group in four dimensions, to which they reduce under compactification on a small two-torus. The geometric modulus of the torus then encodes the coupling constant and theta angle of the four-dimensional theory, and $S$-duality becomes a simple consequence of the invariance under the mapping class group of the torus. Although it is not yet clear exactly what role supersymmetry plays in this context, it seems at least possible that the tensionless string theories, or some other related theories in six dimensions, are the most fundamental theories that do not involve quantum gravity and that they thus `explain' for example the existence of Yang-Mills theories in four dimensions. A proper formulation of tensionless string theories thus appear as a major goal for the future, and it is reasonable to suppose that this is less difficult than understanding quantum gravity theories such as string theory or $M$-theory.

The $A_{N - 1}$ super Yang-Mills theory in four dimensions can be realized as the world-volume theory of $N$ coincident three-branes in type IIB string theory \cite{9510135}. For a single three-brane one gets an abelian vector multiplet consisting of a $U(1)$ gauge field, a set of scalars transforming in the $\bf 6$ representation of the $SO (6) \simeq SU (4)$ $R$-symmetry group, and a set of fermions transforming in the $\bf 4$ representation. When $N$ such branes are brought together, a single vector multiplet representing the `center of mass' collective coordinates decouples, and the remaining $N - 1$ multiplets can be interpreted as spanning the Cartan subalgebra of the $A_{N - 1}$ Lie algebra. Quantization of type IIB strings that stretch between different $D3$-branes furnish $N (N - 1)$ additional vector multiplets that can be interpreted as the roots of $A_{N - 1}$, thus giving rise to the full non-abelian algebra. It is tempting to try to understand the relationship between free $(0, 2)$ tensor multiplets and tensionless string theories in the same way. One would thus consider an $M$-theory configuration where $N$ five-branes are brought together. Factoring out a single `center of mass' tensor multiplet, we are left with $N - 1$ `Cartan algebra' multiplets, and one might think that membranes-branes stretching between different five-branes would supply $N (N - 1)$ additional tensor multiplets spanning the rest of the $A_{N - 1}$ algebra. However, it seems all but impossible to construct a non-abelian version of a two-form (with or without a self-duality constraint on the field strength), and the difficulty here seems to be of a fundamental geometrical nature \cite{9909094}.

The main idea of the present letter is that the proper analogue of a two-form in six dimensions is not a one-form (i.e. an ordinary gauge field) in four dimensions, but rather a zero-form (i.e. a scalar) in two dimensions. A common property of these theories is that they admit a `chiral factorization' (better known as holomorphic factorization in Euclidean signature) \cite{9908107}. In two dimensions, a scalar has a well-understood interacting generalization, namely the Wess-Zumino-Witten model, which is characterized by a positive integer `level' $k$ and an arbitrary Lie group $G$ \cite{Witten84}. These theories also admit holomorphic factorization \cite{Witten92}. The case when $G$ is of $ADE$-type and $k = 1$ admits a realization in terms of rank $G$ free scalars that take their values on the maximal torus of $G$ \cite{Halpern}\cite{Frenkel-Kac}\cite{Segal}. (See e.g. \cite{Lerche-Schellekens-Warner} for a pedagogical review.) The root lattice $\Gamma^r$ and the weight lattice $\Gamma^w$ of the corresponding $ADE$-type Lie algebra play important roles in this construction, which can be naturally generalized to six dimensions.

This surprisingly simple construction reproduces some of the properties of the tensionless string theories. For example, if we consider the $A_{N - 1}$ theory on a six-manifold $M$, we have `Wilson surface' observables $W_x (\Sigma)$, where $x \in {\bf Z}_N$ and $\Sigma$ is a two-cycle in $M$. They obey the selection rule
\be
\left< W_{x_1} (\Sigma_1) \ldots W_{x_p} (\Sigma_p) \right> = 0 \label{selection_rule}
\ee
unless the homology class $[x_1 \Sigma_1 + \ldots x_p \Sigma_p] \in H_2 (M, {\bf Z}_N)$ vanishes \cite{9807205}. Furthermore, the Hilbert space of the theory naturally carries an irreducible representation of the discrete group generated by a set of operators $\Phi_\nu$ where $\nu \in H^3 (M, {\bf Z}_N)$ subject to the relations \cite{9812012}
\be
\Phi_\nu \Phi_{\nu^\prime} = \Phi_{\nu^\prime} \Phi_\nu e^{2 \pi i \nu \cdot \nu^\prime / N} . \label{relations}
\ee
In a situation where we consider the six-dimensional theory on a small two-torus times a four-manifold $Y$, this Hilbert space has a basis labelled by $x \in H^2 (Y, {\bf Z}_N)$, which can be interpreted as the non-abelian magnetic flux of an  $SU (N) / {\bf Z}_N$ Yang-Mills theory  on $Y$ \cite{tHooft}. Under $S$-duality, this flux is related by a Fourier transform to the non-abelian electric flux in the same way as in super Yang-Mills theory, and we also reproduce the geometrical fact that $c_2 + \frac{1}{2} x \cdot x \in H^4 (Y, {\bf Z})$, where $c_2$ and $x$ are the second Chern class and non-abelian magnetic flux of an $SU (N) / {\bf Z}_N$ bundle over $Y$. 

We would like to point out however, that despite these agreements we are still far from the tensionless string theories. Indeed, the number of degrees of freedom of our $A_{N - 1}$ theory only grows linearly with $N$ rather than the $N^3$ growth of the corresponding tensionless string theory \cite{9708005}\cite{9604089}\cite{9806087}, so we are therefore at most getting a subsector of the latter. It is also quite possible that our models have no relationship at all to the tensionless string theories. But they seem to be consistent theories with a rich structure, and the study of their two-dimensional analogues has certainly been fruitful, so given the scarcity of higher-dimensional conformal field theories, we think that they do deserve further study on their own.

\section{The quantum theory of a chiral two-form}
On a six-manifold $M$, we consider a two-form `gauge field', locally given by an ordinary two-form $X$ subject to the gauge equivalence relation $X \sim X + \Delta X$, where $\Delta X$ is a closed two-form with integer periods. The field strength $H = d X$ of $X$ is a closed three-form with integer periods. Given an integer $k$ and a two-cycle $\Sigma$ in $M$, we can construct a `Wilson surface' observable $W_k (\Sigma)$ as $W_k (\Sigma) = \exp 2 \pi i k \int_\Sigma X$. The integrality of $k$ is necessary to ensure gauge invariance of $W_k (\Sigma)$. It is natural to couple $X$ to a background three-form $C$ subject to the gauge equivalence relation $C \sim C + \Delta C$, where $\Delta C$ is a closed three-form with integer periods. 

We now endow $M$ with a conformal structure $[g]$, i.e. a conformal equivalence class of metrics $g$ on $M$. (Formally these theories are conformally invariant, and in this letter we will disregard the Weyl anomalies that afflict the partition function \cite{0001041} and the correlation functions \cite{9905163}.) The conformal structure $[g]$ gives rise to a Hodge duality operator $*$, which is a map from the space of three-forms on $M$ to itself. If $[g]$ has Minkowski signature, we have $* * = 1$ so an arbitrary three-form can be decomposed into its self-dual and anti self-dual parts. However, for a generic conformal structure $[g]$ on $M$, it does not make sense in classical field theory to impose the self-duality condition $* H = H$ on the field strength, since this equation is incompatible with the integrality condition on the periods of $H$. In fact, the self-duality condition does not follow as an equation of motion from a Lagrangian in the usual sense. Nevertheless, it is possible to formulate the quantum theory of a `chiral' two-form (i.e. with self-dual field strength), as we will now review \cite{9610234}.

The background three-form $C$ is a priori an arbitrary three-form on $M$, but certain symmetries allow us to restrict it be an element of the space $\Omega^0$ of harmonic three-forms on $M$. The space of gauge-inequivalent such $C$-fields is the torus $\Omega^0 / \Delta$ (the intermediate Jacobian of $M$), where $\Delta$ is the lattice of harmonic three-forms on $M$ with integer periods. The wedge product $\wedge$ followed by integration over $M$ endows $\Omega^0 / \Delta$ with a symplectic structure, and we denote the corresponding symplectic form as $\omega$. We have $\int_{\Omega^0 / \Delta} \omega^{\frac{1}{2} b_3} / (\frac{1}{2} b_3)! = 1$, so $\omega$ gives $\Omega^0 / \Delta$ the structure of a principally polarized abelian variety. The partition function of the chiral two-form $X$ is a section of a line-bundle ${\cal L}$ over $\Omega^0 / \Delta$, the curvature of which equals $\omega$.
 
We will now temporarily assume that $[g]$ instead has a Euclidean signature. The duality operator then obeys $* * = -1$, and thus gives $\Omega^0 / \Delta$ the structure of a complex torus of complex dimension $\frac{1}{2} b_3 (M)$. The moduli space ${\cal J}$ of {\it all} complex structures $J$ on  $\Omega^0 / \Delta$ (i.e. not restricting to those that arise from a conformal structure $[g]$ on $M$) is itself a complex manifold. The symplectic form $\omega$ is of type $(1, 1)$ with respect to $J$, so ${\cal L}$ is in fact a holomorphic line-bundle. (In applications, for example to the five-brane in $M$-theory, the physical context provides additional data that completely determines ${\cal L}$.) The partition function of $X$ can be seen to depend holomorphically on the complex structure $J \in {\cal J}$ and to be a holomorphic section of ${\cal L}$. It follows from the Riemann-Roch theorem together with the Kodaira vanishing theorem that if ${\cal M}$ is a holomorphic line-bundle over $\Omega^0 / \Delta$ with $c_1 ({\cal M}) = [N \omega]$ for some positive integer $N$, then the space $H^0 (\Omega^0 / \Delta, {\cal M})$ of holomorphic sections of ${\cal M}$ has dimension $\dim H^0 (\Omega^0 / \Delta, {\cal M}) = N^{\frac{1}{2} b_3}$. (See e.g. chapter 2, section 6 of \cite{Griffiths-Harris}.) The line-bundle ${\cal L}$, for which $N = 1$, thus has a (up to a multiplicative constant) unique holomorphic section, known as the Jacobi theta function. 

To give a more explicit description of the above construction, we identify $\Omega^0$ with ${\bf C}^{\frac{1}{2} b_3}$ and choose coordinates so that $\Delta$ is spanned by the vectors $e^{(1)} = (1, 0, \ldots, 0), \ldots, e^{(\frac{1}{2} b_3)} = (0, \ldots, 0, 1)$ and the vectors $Z^0 e^{(1)}, \ldots, Z^0 e^{(\frac{1}{2} b_3)}$. Here the so called period matrix $Z^0 = (Z^0_{i j})$, $i, j = 1, \ldots, \frac{1}{2} b_3$ parametrizes the complex structure of $\Omega^0 / \Delta$ and takes its values in a generalized upper halfplane, i.e. $Z^0 = {}^t Z^0$ and ${\rm Im} \, Z^0 > 0$. With the background three-form given by $C = (C_1, \ldots, C_{\frac{1}{2} b_3}) \in {\bf C}^{\frac{1}{2} b_3} / \Delta$, the Jacobi theta function is given by
\be
\theta (Z^0 | C) = \sum_{k_i \in {\bf Z}} e^{\pi i k_i Z^0_{i j} k_j + 2 \pi i k_i C_i} .
\ee
(For simplicity, we have considered a line bundle ${\cal L}$ with `zero characteristics'.)

Actually, the partition function of the chiral two-form theory is not quite given by the above formula. We have used a trivialization of ${\cal L}$ in which the transition functions are holomorphic and ${\bf C}^*$-valued. Instead, one should really use a trivialization with $U (1)$-valued transition functions. This amounts to multiplying the Jacobi theta function with a certain prefactor \cite{9908107}. Furthermore, the theta function actually describes the contribution to the partition function from `on-shell' field configurations, and strictly speaking one should also take quantum fluctuations into account \cite{9806016}\cite{9908107}. These will not play any role for the topological effects discussed in this letter, though, and we will therefore disregard them. One could also take the point of view that for the complete supersymmetric tensor multiplet, the quantum fluctuations of the two-form should cancel against those of the scalars and the spinors.

\section{The theory of multiple two-forms}
We now consider $N$ chiral two-forms $X^I$, $I = 1, \ldots, N$ on a six-manifold $M$. If we assume that they are non-interacting, their partition function would just be the product of the partition functions of the separate two-forms. If we also assume that they all couple to the same background three-form $C$, as is appropriate for the case of $N$ coincident five-branes in $M$-theory, the partition function for the whole system is thus
\be
\left(\theta (Z^0 | C) \right)^N = \sum_{\bar{k}_i \in {\bf Z}^N} e^{\pi i \bar{k}_i \cdot Z^0_{i j} \bar{k}_j + 2 \pi i \bar{k}_i \cdot \bar{1} C_i} ,
\ee 
where $\bar{1} = (1, \ldots, 1) \in {\bf Z}^N$ and the raised dot denotes the standard inner product between elements of ${\bf R}^N$. 

In analogy with the case of $N$ coincident three-branes in type IIB string theory, where the gauge group of the world-volume theory is $SU (N)$ rather than $U (N)$, we wish to factor out the contribution from the collective `center of mass' degrees of freedom. We therefore decompose the summation variable $\bar{k}_i \in {\bf Z}^N$ uniquely as $\bar{k}_i = (x_i + n_i) \bar{1} + \bar{w}_i$, where $x_i \in \{ 0, \frac{1}{N}, \ldots, \frac{N - 1}{N} \}$, $n_i \in {\bf Z}$ and $\bar{w}_i \in \Gamma_{x_i}$. Here
\be
\Gamma_{x_i} = \{ \bar{w}_i \in (- x_i + {\bf Z})^N; \bar{w}_i \cdot \bar{1} = 0 \} . \label{Gamma_x} 
\ee
We can then rewrite the partition function as
\be
\left(\theta (Z^0 | C) \right)^N = \sum_{x_i \in \{ 0, \frac{1}{N}, \ldots, \frac{N - 1}{N} \}} s_x (Z^0 | C) \theta_x (Z^0) ,
\ee
where
\be
s_x (Z^0 | C) = e^{N \pi i x_i Z^0_{i j} x_j + 2 N \pi i x_i C_i} \sum_{n_i \in {\bf Z}} e^{N \pi i n_i Z^0_{i j} n_j + 2 N \pi i n_i (C_i + Z^0_{i j} x_j)}
\ee
and
\be
\theta_x (Z^0) = \sum_{\bar{w}_i \in \Gamma_{x_i}} e^{\pi i \bar{w}_i Z^0_{i j} \bar{w}_j} . \label{theta_x}
\ee

The $N^{\frac{1}{2} b_3}$ different functions $s_x (Z^0 | C)$ have the same transition functions, and in fact span the space $H^0 (\Omega^0 / \Delta, {\cal L}^N)$ of holomorphic sections of the line bundle ${\cal L}^N$. This bundle is invariant under translations of $\Omega^0 / \Delta$ by vectors of the form $\alpha_i e^{(i)} + \beta_i Z^0 e^{(i)}$, where $\alpha_i, \beta_i \in \{0, \frac{1}{N}, \ldots, \frac{N - 1}{N} \}$. We thus have a set of translation operators $\Phi_\nu$ indexed by $\nu = (\alpha_1, \ldots, \alpha_{\frac{1}{2} b_3}; \beta_1, \ldots, \beta_{\frac{1}{2} b_3})$ and acting on a section $s (Z^0 | C)$ of ${\cal L}^N$ as
\be
\left(\Phi_\nu s \right) (Z^0 | C) = e^{2 N \pi i \beta_i C_i} s (Z^0 | C + \alpha_i e^{(i)} + \beta_i Z^0 e^{(i)}) .
\ee
These operators fulfil the relations (\ref{relations}) with $\nu \cdot \nu^\prime / N = N (\beta^\prime_i \alpha_i - \beta_i \alpha^\prime_i)$. We interpret the sections $s_x (Z^0 | C)$ as being the contributions to the partition function from the collective coordinates, whereas the functions $\theta_x (Z^0)$ form a vector of different partition functions for the `internal' degrees of freedom. This internal theory is of course our real object of interest.  

The theory described above can be regarded as the $A_{N - 1}$ version of a larger class of theories with an $ADE$-classification. To see this, we start by considering the root lattice $\Gamma^r$ of the $A_{N - 1}$ Lie algebra, i.e. $su (N)$. This is of rank $N - 1$, but is most conveniently described in $N$ dimensions:
\be
\Gamma^r = \{ \bar{r} = (r_1, \ldots, r_N) \in {\bf Z}^N; \bar{r} \cdot \bar{1} = 0 \} .
\ee
Being an integral (in fact even) lattice, $\Gamma^r$ is a sublattice of its dual $\Gamma^w$, which is the weight lattice of $su (N)$:
\be
\Gamma^w = \{ \bar{w} = (w_1, \ldots, w_N) \in (\frac{1}{N} {\bf Z})^N; w_a - w_b \in {\bf Z}, \bar{w} \cdot \bar{1} = 0 \} .
\ee
We can therefore decompose $\Gamma^w$ in cosets with respect to $\Gamma^r$. The quotient group $\Gamma^w / \Gamma^r$ is isomorphic to the center ${\bf Z}_N$ of the simply connected Lie group $SU (N)$. The coset $\Gamma_{x_i}$ for $x_i \in \{0, \frac{1}{N}, \ldots, \frac{N - 1}{N} \}$ is in fact defined in (\ref{Gamma_x}). The function $\theta_x (Z^0)$ defined in (\ref{theta_x}) is known as a conjugacy class theta function (see e.g. \cite{Lerche-Schellekens-Warner}). The selection rule (\ref{selection_rule}) is a generalization to surface observables and six dimensions of the fact that correlation functions of point observables in two-dimensional Wess-Zumino-Witten models vanish unless the total conjugacy class adds up to zero. These considerations are immediately generalizable to the $D$- and $E$-series.  

\section{Relationship to $N = 4$ super Yang-Mills theory in four dimensions}
Finally, we will consider the special case when the six-manifold $M$ is of the form $T2 \times Y$ with $T2$ a two-torus of modulus $\tau = \tau_1 + i \tau_2$ and $Y$ a four-manifold. We then have $H^3 (M, {\bf Z}_N) \simeq H^2 (Y, {\bf Z}_N) \oplus H^2 (Y, {\bf Z}_N)$, and the $x$ that labels the conjugacy class theta functions can be viewed as an element of $H^2 (Y, {\bf Z}_N)$. With respect to this basis, we have $Z^0_{ij} = \tau_1 S_{ij} + i \tau_2 T_{ij}$, where $S$ is the intersection form and $S^{-1} T$ describes the action of the Hodge duality operator $*$ on $H^2 (Y, {\bf Z})$. This period matrix obeys $-(Z^0)^{-1} = \tau_1^\prime S^{-1} + i \tau_2^\prime S^{-1} T S^{-1}$, where $\tau^\prime = \tau_1^\prime + i \tau_2^\prime = - 1 / \tau$, and $\det Z^0 = \tau^{b_2^- (Y)} \bar{\tau}^{b_2^+ (Y)}$, where $b_2^+ (Y)$ and $b_2^- (Y)$ are the dimensions of the self-dual and anti self-dual parts of $H^2 (Y, {\bf R})$ respectively. Writing $\theta_x (\tau, S, T)$ for $\theta_x (\tau_1 S + i \tau_2 T)$, it now follows from the modular transformation properties of the conjugacy class theta functions that
\bea
\theta_x (\tau + 1, S, T) & = & e^{2 \pi i \frac{1}{2} x_i S_{i j} x_j (N^2 - N)} \theta_x (\tau, S, T) \cr
\theta_x (-1 / \tau, S^{-1}, S^{-1} T S^{-1}) & = & \frac{1}{\sqrt{N}} \tau^{\frac{N - 1}{2} b_2^- (Y)} \bar{\tau}^{\frac{N - 1}{2} b_2^+ (Y)} \sum_{x^\prime} e^{2 N \pi i x_i S_{ij} x_j^\prime} \theta_{x^\prime} (\tau, S, T) . 
\eea

 To make contact with $SU (N) / {\bf Z}_N$ super Yang-Mills theory on $Y$, we interpret $\tau$ as the combination $\frac{\theta}{2 \pi} + \frac{4 \pi i}{g^2}$ of the coupling constant $g$ and the theta angle $\theta$, and $\theta_x (\tau, S, T)$ and $\theta_x (\tau, S^{-1}, S^{-1} T S^{-1})$ as the partition functions of the theory in the sector with non-abelian magnetic or electric flux $x \in H^2 (Y, {\bf Z}_N)$ respectively. The above formulas then agree with the behaviour of super Yang-Mills theory under $S$-duality: The phase that is picked up under the transformation $\tau \rightarrow \tau + 1$, i.e. under a shift of the theta angle by $2 \pi$, correctly reflects the non-integrality of the second Chern-class for a bundle with non-zero non-abelian magnetic flux, and the transformation $\tau \rightarrow -1 / \tau$ is accompanied by an exchange of the magnetic and electric fluxes, which are discrete Fourier transforms of each other \cite{9408074}. The modular weights generalize those that appear in \cite{9505186}.

\end{document}